\documentstyle[11pt,paspconf,epsfig]{article}

\begin{document}

\title{Lensing by Groups of Galaxies}
\author{Ole M\"oller}
\affil{Cavendish Laboratory, Madingley Road, Cambridge CB3 0HE, UK}
\author{Priyamvada Natarajan}
\affil{Institute of Astronomy, Madingley Road, Cambridge CB3 0HA, UK}
\begin{abstract}
A large fraction of known galaxy-lens
systems require a component of external shear to explain the observed
image geometries. In most cases, this shear can be attributed
to a nearby group of galaxies. We discuss how the dark-matter mass distribution of groups 
of galaxies can influence the external shear for strongly lensed sources and  
calculate the expected weak lensing signal from groups for various mass profiles.
\end{abstract}
\keywords{gravitational lensing, galaxies: fundamental parameters, halos, methods: numerical}
Most galaxies in the universe are neither in clusters nor isolated,
but found in groups.  To date, it is not known whether there exists a
significant group dark-matter halo, similar to that for clusters of
galaxies, or whether most of the matter is associated with the
individual galaxies themselves. A recent weak-lensing study of a
number of groups has shown that it is possible in principle to
determine the mass profile of groups using their weak lensing signal
(Hoekstra 1999). Because many models of strong lens systems require a
substantial external shear to explain the observed image geometries
and magnifications (Keeton, Kochanek \& Seljak 1997, Kneib et al. 1998), we investigate
how  the dark matter distribution in a nearby group affects the
external shear. We outline how both strong and weak
gravitational lensing can be used to  determine the mass distribution
within groups of galaxies.

\begin{figure*}
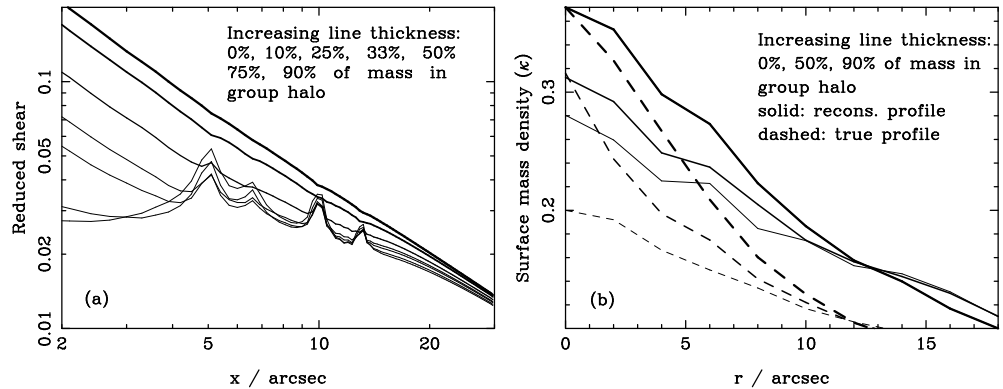

\label{figures}
\begin{center}
\vskip -5mm \epsfig{file=omoeller1.ps,width=5.1cm,angle=-90} \hskip 1mm
\epsfig{file=omoeller2.ps,width=5.1cm,angle=-90}\\
\end{center}
\caption{The effect of the mass distribution model on the lensing signal for
various values of the group halo mass fraction. (a) The reduced shear as a function of radial distance from the group centre. 
(b) The weak lensing mass reconstruction from
simulated observations and the theoretical profiles for three values of the group halo mass fraction.}
\end{figure*}
To investigate the gravitational lensing effect of groups we use a
fast and accurate numerical ray-tracing code (M\"oller \& Blain 1998, Blain, M\"oller \& Maller 1999) 
and a group model
based on PG115+080 (Keeton \& Kochanek 1997).  The lens PG115+080 consists of four galaxies  separated by
$\approx10''$ with a total velocity dispersion of about
$270\,\mathrm{km\,s^{-1}}$.  The group halo and 
galaxies are modelled as singular isothermal spheres
(SISs). The group halo is assumed to be centred on the geometrical mean position of
the group galaxies. We carried out simulations for different values of the
group halo : individual galaxy mass ratio. Fig.\,1(a) 
shows the expected radial profile of the reduced shear. There is a significant
difference between a model in which all the mass is associated with the
galaxies and models in which a significant fraction of the mass is in
an intergalactic group halo. This will affect both the weak lensing
distortions of background galaxies and the external shear
contribution to any nearby strong galaxy lens.  To test
whether weak lensing observations can be used to distinguish these two
different cases we simulated 100 $60''\times60''$ fields at a 
resolution of $0.1''$, each containing 500
galaxies. The future Advanced Camera for Surveys (ACS) on the Hubble Space Telescope will have a field of view of similar size, a resolution slightly below $0.1''$ and a sensitivity that will yield several hundreds of galaxies per square arc-minute in a few hours of integration. Fig.\,1(b) shows the reconstructed mass profiles. For groups that have a massive group halo the mass profile is significantly steeper than for groups in which most mass is associated with the galaxies. The offset between the reconstructed and true mass profiles is due to the
mass sheet degeneracy. The results show that it would be possible to distinguish between the different mass models with observations of this quality, which will be practical using ACS.

The external shear produced by a galaxy group depends
significantly on the mass profile of the group. As most galaxies, and
hence lenses, are likely to be part of a group, most
lens models will need to take this effect into account.  In the
weak lensing regime, our results show that observations with new
instruments, like the ACS, could differentiate between various group mass
profiles.
\acknowledgments
We thank Andrew Blain for useful discussions and comments on the manuscript.

\end{document}